\documentclass[prc,superscriptaddress,unsortedaddress,twocolumn,showpacs,preprintnumbers,amsmath,amssymb]{revtex4}
\usepackage[dvips]{graphicx}
\usepackage{dcolumn}
\usepackage{amsmath}

\begin{document}

\title{ New treatment of breakup continuum \\
in the method of continuum discretized coupled channels}

\author{T. Matsumoto}
\email[Electronic address: ]{taku2scp@mbox.nc.kyushu-u.ac.jp}
\affiliation{Department of Physics, Kyushu University, Fukuoka 812-8581,
Japan}
\author{T. Kamizato}
\affiliation{Department of Physics and Earth Sciences,
University of the Ryukyus, Nishihara-cho, Okinawa 903-0213, Japan}
\author{K. Ogata}
\affiliation{Department of Physics, Kyushu University, Fukuoka 812-8581,
Japan}
\author{Y. Iseri}
\affiliation{Department of Physics, Chiba-Keizai College, Todoroki-cho
4-3-30, Inage, Chiba 263-0021, Japan}
\author{E. Hiyama}
\affiliation{Institute of Particle and Nuclear Studies, High Energy
Accelerator Research Organization (KEK), Tsukuba, 305-0801, Japan}
\author{M. Kamimura}
\affiliation{Department of Physics, Kyushu University, Fukuoka 812-8581,
Japan}
\author{M. Yahiro}
\affiliation{Department of Physics and Earth Sciences,
University of the Ryukyus, Nishihara-cho, Okinawa 903-0213, Japan}

\date{\today}

\begin{abstract}
 A new method of pseudo-state discretization is proposed for the method
 of continuum discretized coupled channels (CDCC) to deal with
 three-body breakup processes. We propose real- and
 complex-range Gaussian bases for the pseudo-state wave
 functions, and show that they form in good approximation a complete set 
in the configuration space which is important for breakup processes. 
Continuous S-matrix elements are derived 
with the approximate completeness from discrete ones calculated by CDCC. 
 Accuracy of the method is tested quantitatively for two realistic
 examples, $d$+$^{58}$Ni scattering at 80 MeV and $^{6}$Li+$^{40}$Ca
 scattering at 156 MeV with the satisfactory results. Possibility of
 application of the method to four-body breakup processes is also discussed.

\end{abstract}


\maketitle

\section{Introduction}
\label{sec:introduction}

The method of continuum discretized coupled channels (CDCC)
has been successful in describing nuclear reactions including weakly bound
projectiles~\cite{CDCC-review1,CDCC-review2,Yahiro1,yahiro2,sakuragi1,%
sakuragi2,sakuragi3,sakuragi4,iseri1,Sakuragi,Surrey,Ogata}.
CDCC has been attracting much attention since
the advent of experiments with radioactive beams, because 
projectile breakup processes are essential to many of such reactions.
CDCC plays an important role in the spectroscopic studies of radioactive
nuclei through the nuclear reactions involving such nuclei.

In CDCC for reactions with a projectile consisting of two fragments,
the states of the projectile are classified by the linear and the
angular momenta, $k$ and $\ell$, 
of relative motion of the two fragments of the projectile, 
which are truncated by
$k \le k_{\rm max}$ and $ \ell \le \ell_{\rm max}$.
The truncation is the most basic assumption in CDCC, and
it is confirmed that
calculated $S$-matrix elements converge
for sufficiently large $k_{\rm max}$ and $\ell_{\rm max}$%
~\cite{CDCC-review1,Yahiro1,Piya}. 
It has been shown 
that CDCC is the first-order approximation to the distorted
Faddeev equations,
and corrections to the converged CDCC solution are
negligible within the region of space in which the reaction takes place%
~\cite{CDCC-foundation}. 

As a consequence of the truncation, the 
integral equation form of the equations of coupled channels,
derived from the three-body Schr\"{o}dinger equation,
has a compact kernel,
satisfying a necessary condition for iterative solutions%
~\cite{CDCC-foundation}.
In practice, however, 
the coupled channels equations thus obtained are impossible to be solved
because of the continuously infinite number of coupled breakup channels.
The problem is solved by discretizing the $k$ continuum.
The discretization leads the 
coupled equations to a set of differential equations 
with a finite number of channels.

As for the discretization, three methods have been proposed so far:
the average (Av)~\cite{YK1,Yahiro1,CDCC-review1,CDCC-review2},
the midpoint (Mid)~\cite{CDCC-review2, Piya}, and
the pseudo-state (PS)~\cite{CDCC-review1,PS1, PS2} methods.
In the Av and Mid methods, the
$k$-continuum is divided into a finite number of bins.
In the former the continuum channels within each bin are
averaged into a single channel, while in the latter they are
represented by the channel at a midpoint of the bin.
It has been confirmed 
that calculated $S$-matrix elements converge
as the width $\Delta$ of the bins is decreased,
and also that the two methods yield the same converged $S$-matrix
elements~\cite{CDCC-review1,Yahiro1,Piya}. 
From practical point of
view, the Av method is more convenient than the Mid one,
since the former requires less numerical works than the latter.
The Av method therefore is most widely used in practice.

In the PS method, the breakup states are described by superpositions
of $L^2$-type basis functions.
The wave functions of such pseudo breakup state have wrong 
asymptotic forms. For this reason, 
the PS method was mainly used in the past to describe 
virtual breakup processes in the
intermediate stage of elastic scattering and ($d,p$) reactions. Very
recently, the applicability of the PS method for real breakup reactions
for a specific set of basis functions.
It was shown that the PS method well simulates the 
angular distribution of the total breakup cross section calculated with
the Av method ~\cite{PS1}. 
In the analysis, discrete breakup cross sections calculated with 
the PS method are transformed into continuous one, 
by assuming a specific form for the $k$ distribution 
of the continuous cross section, that is, a histogram with
widths estimated in a reasonable way.

The purpose of the present paper is to propose a new PS method for 
projectile breakup reactions. 
If the basis functions form in the good approximation a complete set in 
the configuration space which is important for breakup processes, 
we can formulate a method of interpolation for generating 
continuous breakup $S$-matrix elements from the discrete ones provided by 
the PS method, following the formulation of \cite{Tostevin} for the Av method 
in which the completeness of discretized states is well satisfied
if the discretization is enough precise. 
The method does not assume any form a priori 
for $k$ distributions of continuous breakup $S$-matrix elements. 
The method is independent of the type of the basis functions 
used for the PS method. 
Only condition is that the basis 
functions constitute an approximate complete set 
in the wide range of
$k$ and its conjugate coordinate $r$ 
which are important for the breakup processes. 
As basis functions satisfying this condition, we propose two types of 
them. One is ordinary Gaussian functions~\cite{K-G}, which we refer to
as {\it real-range Gaussian function} in the present paper.
The other is a natural extension of that,
{\it complex-range Gaussian function}~\cite{H-Ka-Ki},
i.e, Gaussian functions with the complex range parameters 
the precise definition of which is given in a later section.

In Section II, we recapitulate CDCC based on both the Av and the
PS methods of discretization.
In Section III, we present a method of interpolation to get continuous
breakup $S$-matrix elements from discrete ones calculated
with the PS method, and introduce the real and the complex-range
Gaussian basis.
In Section IV, the validity of the present PS method is tested and
justified for two realistic cases, $d$+$^{58}$Ni scattering at 80 MeV
and $^{6}$Li+$^{40}$Ca scattering at 156 MeV.
In Section V, the potentiality of the present PS method for
four-body breakup reactions is discussed.
Section VI gives a summary.

\section{The Method of Continuum Discretized Coupled Channels}
\label{sec:CDCC-method}

We consider a reaction of a weakly bound projectile (B) impinging on a
target nucleus (A). 
We treat a simple system shown in Fig.~1 
in which the projectile, B, is composed of two particles (b and c) and
the target is inert. The three-body system is described
by a model Hamiltonian
$H = H_{\rm bc} + K_{R} + U ,$
where $H_{\rm bc}=K_{r}+V_{\rm bc}({\bf r})$ and
$U=U_{\rm bA}({\bf r}_{\rm bA}) + U_{\rm cA}({\bf r}_{\rm cA})$.
Vector ${\bf r}$ is the relative
coordinate between b and c, ${\bf R}$ 
the one between the center-of-mass of the b-c pair
and the center-of-mass of A,
and ${\bf r}_{\mathrm{XY}}$ 
denotes the relative coordinate between two particles X and Y.
Operators $K_{r}$ and $K_{R}$ are kinetic energies
associated with ${\bf r}$ and ${\bf R}$, respectively.
$V_{\rm bc}({\bf r})$ is the interaction between b and c.
The interaction
$U_{\rm bA}$ ($U_{\rm cA}$) between b (c) and A
is taken to be
the optical potential for b+A (c+A) scattering.
For simplicity, the spin dependence of the interactions is neglected.
Furthermore, the Coulomb part of $U$ is
treated approximately as a function only of ${\bf R}$,
i.e., we neglect Coulomb breakup processes,
and focus our attention on nuclear breakup.

\begin{figure}[htbp]
\begin{center}
 \includegraphics[width=0.26\textwidth,clip]{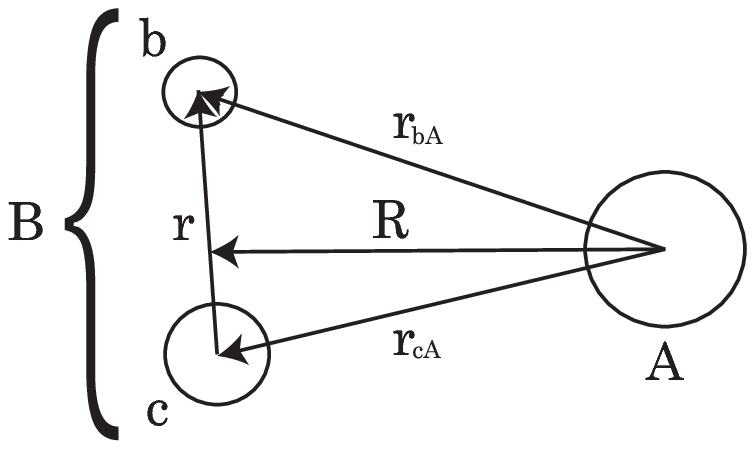}
 \caption{Illustration of a three-body (A+b+c) system.
 The symbol B=b+c stands for the projectile, and A is the target.
}
\end{center}
\end{figure}

In CDCC, the three-body wave function $\Psi_{JM}$,
with the total angular momentum $J$
and its projection $M$ on $z$-axis, is expanded
in terms of the orthonormal set of eigenstates $\Phi$ of
$H_{\rm bc}$:
\begin{eqnarray}
&&\Psi_{JM} ({\bf r,R})
=\sum_{L}
{\cal Y}_{JM}^{\ell_{0},L}
\Phi_{0}(r)
\chi_{\ell_0 LJ}(P_0,R)/R
\nonumber \\
&&+
\sum_{\ell,L}
{\cal Y}_{JM}^{\ell,L}
\int_{0}^{\infty}
\Phi_{\ell}(k,r)
\chi_{\ell LJ}(P,R)/R \,\;dk ,
\label{exact-expansion}
\end{eqnarray}
where
\begin{equation}
{\cal Y}_{JM}^{\ell,L}=
[i^{\ell }Y_{\ell}(\Omega_r)\otimes i^L Y_{L}(\Omega_R)]_{JM}.
\end{equation}
For simplicity, we assume that the b+c system has one bound state
$\Phi_{0}(r)$ with angular momentum $\ell_0$ and
continuum states $\Phi_{\ell}(k,r)$
with linear momentum $k$ and angular momentum $\ell$,
both ranging from zero to infinity.
The $\Phi_{\ell}$ are real functions normalized
to the $\delta$-function in $k$~\cite{Yahiro1}.
The projectile B is initially in the bound state.
The coefficient  $\chi_{\ell LJ}$ ($\chi_{\ell_0 LJ}$)
of the expansion
describes a center-of-mass motion of the b-c pair
relative to A in the state $\Phi_{\ell}$ 
with the linear and orbital angular momenta
$P$ $(P_0)$ and $L$, respectively.

In CDCC, the sum over $\ell$
is truncated by $\ell \leq \ell_{\rm max}$ and the $k$ integral
by $k \leq k_{\rm max}$.
For each $\ell$, furthermore, the continuum states from $k=0$ to
$k_{\rm max}$ are discretized into a finite number of states,
with the wave function $\hat{\Phi}_{i\ell}(r)$ of the i-th state
corresponding to the momentum $\hat{k}_i$. 
Details of the discretization methods are described in the next
section.

After the truncation and the discretization,
$\Psi_{JM}$ is reduced to an approximate one,
\begin{eqnarray}
 &&\Psi^{\rm CDCC}_{JM} =
 \sum_{L}
 {\cal Y}_{JM}^{\ell_{0},L}
 \Phi_{0}(r)
 \hat{\chi}_{\gamma_0}(P_0,R)/R
 \nonumber \\
&& +
 \sum_{l=0}^{l_{\rm max}}  \sum_{i=1}^{N} \sum_{L}
{\cal Y}_{JM}^{\ell,L}\hat {\Phi}_{i\ell} (r)
\hat{\chi}_{\gamma}(\hat{P}_i,R)/R,
\label{appro-expansion}
\end{eqnarray}
where
\begin{eqnarray}
\hat{\chi}_{\gamma_0}(P_0,R) &=& \chi_{\gamma_0}(P_0,R),
\;\;\;
\gamma_0 = (0,\ell_0,L,J),
\nonumber
\\
\hat{\chi}_{\gamma}(\hat{P}_i,R) &=&
W_{\gamma} \chi_{\gamma}(\hat{P}_i,R),
\;\;\;
\gamma = (i,\ell,L,J).
\nonumber
\label{relation}
\end{eqnarray}
On the right hand side of Eq.~(\ref{appro-expansion}), the first term
represents the elastic channel denoted by
$\gamma_0$
and the second one corresponds to the discretized breakup channels, each
denoted by $\gamma$.
The weight factor $W_{\gamma}$ depends on the discretization method 
used. Each pair of momenta, $(\hat{k}_i, \hat{P}_i)$,
satisfies the total energy conservation:
\begin{equation}
 E=\hbar^{2} P_{0}^{2}/2\mu_{\rm AB}+\epsilon_0=
\hbar^{2} \hat{P}_{i}^{2}/2\mu_{\rm AB}+\epsilon_i ,
\end{equation}
where
$\epsilon_0$ and $\epsilon_i=\hbar^{2} \hat{k}_{i}^{2}/2\mu_{\rm bc}$
are the energies of the ground and the continuum states, respectively.
Inserting Eq.~(\ref{appro-expansion}) into
the three-body Schr\"{o}dinger equation, $(H-E)\Psi_{JM}=0$, leads to
a set of coupled differential equations for
$\hat{\chi}_{\gamma_0}(P_0,R)$ and $\hat{\chi}_{\gamma}(\hat{P}_i,R)$:
\begin{eqnarray}
   &&\left[\frac{d^{2}}{dR^{2}} + \hat{P}_{i}^{2} - \frac{L(L+1)}{R^{2}}
 - \frac{2 \mu_{\rm AB}}{\hbar^{2}} V_{\gamma \gamma}(R)\right]
\hat{\chi}_{\gamma}(\hat{P}_i,R)
\nonumber \\
 && = \sum_{\gamma^{'}\neq\gamma} \frac{2 \mu_{\rm AB}}{\hbar^{2}}
   V_{\gamma \gamma ^{'}}(R) \hat{\chi}_{\gamma^{'}} (\hat{P}_{i^{'}},R)
\label{eqs:CDCC}
\end{eqnarray}
for all $\gamma$ including $\gamma_0$,
where $\hat{k}_0=\sqrt{-2\mu_{\rm bc}\epsilon_0}/\hbar$ and
$\hat{P}_0=P_0$. 
The coupling potentials $V_{\gamma\gamma^{'}}(R)$
are obtained as
\begin{equation}
  V_{\gamma \gamma^{'}}(R)   =
     \langle {\cal Y}_{JM}^{\ell,L}\hat {\Phi}_{i\ell} (r)
     | U | {\cal Y}_{JM}^{\ell^{'},L^{'}}\hat {\Phi}_{i^{'}\ell^{'}} (r)
     \rangle_{{\bf r}, \Omega_R}.
\label{coupling}
\end{equation}
The coupled equations are solved,
under the asymptotic boundary condition
\begin{equation}
 \hat{\chi}_{\gamma}(\hat{P}_i,R) \sim u_{L}^{(-)} (\hat{P}_{i},R)
\delta_{\gamma , \gamma_0} - \sqrt{\frac{\hat{P}_i}{\hat{P}_{0}}}
\hat{S}_{\gamma , \gamma_0} u_{L}^{(+)} ( \hat{P}_{i} , R).
 \end{equation}
Here
$u_{L}^{(-)}(\hat{P}_{i},R)$ and $u_{L}^{(+)}(\hat{P}_{i},R)$
are incoming and outgoing Coulomb wave functions
with the momentum $\hat{P}_{i}$,
and $\hat{S}_{\gamma,\gamma_0}$ is the $S$-matrix
element for the transition from the initial channel $\gamma_0$
to $\gamma$.

\section{Discretization of $k$ continuum}
\label{sec:discretization}

Among the three methods of discretization of the $k$ continuum,
the relation between the average (Av) and the midpoint (Mid) methods
has already been clarified~\cite{Piya}.
The present discussion therefore is focused on the relation between
the Av and the pseudo-state (PS) methods.

\subsection{The average method }
\label{sec:Av-method}

In the Av method, the $k$-continuum [0, $k_{\rm max}$], for each $\ell$,
is divided into a finite number of bins,
each with a width $\Delta_{i\ell}=k_{i}-k_{i-1}$, and
the continuum breakup states in the $i$-th bin
are averaged with a weight function
$f_{i\ell}(k)$~\cite{CDCC-review1,CDCC-review2}.
The resultant orthonormal state is described as
\begin{equation}
\label{state-AV}
 \hat{\Phi}_{i\ell}(r) =
 \frac{1}{W_{\gamma}} \int_{k_{i-1}}^{k_{i}}
\Phi_{\ell}(k, r) f_{i\ell}(k) dk
\quad
 {\mbox{(for Av) ,}}
\end{equation}
then the weight factor $W_\gamma$ is given by
\begin{equation}
   W_{\gamma}^2=\int_{k_{i-1}}^{k_{i}}[f_{i\ell}(k)]^2\, dk.
\nonumber
\end{equation}
For a bin far from a resonance,
it is natural to set
$f_{i\ell}(k)=1$, so that $W_{\gamma}=\sqrt{\Delta_{i\ell}}$ 
since $\Phi_{\ell}(k,r)$ changes smoothly with $k$.
On the other hand, $\Phi_{\ell}(k,r)$ changes rapidly across the
resonance. One way of coping with this situation is to take
$\Delta_{i\ell}$ much smaller than the width of the resonance so that
$\Phi_{\ell}(k,r)$ does not change much within individual bins. This,
however, make the number of bins large. Alternatively, one can take a
single bin which contains the whole resonance peak and
use a weight function of Breit-Wigner
type~\cite{CDCC-review1,sakuragi1,sakuragi2,sakuragi4,Sakuragi},
\begin{equation}
f_{i\ell}(k)=\left|\frac{i\Gamma/2}
               {\epsilon(k)-\epsilon_{\rm res}+i\Gamma/2}\right| ,
\label{BW-weight}
\end{equation}
where $\epsilon(k)$ is a continuous intrinsic energy of the b+c system.
The discretized intrinsic energy,
$\epsilon_i=\hbar^{2} \hat{k}_{i}^2/2\mu_{\rm bc}$,
corresponding to  each bin is obtained as
$\hat{k}_{i}^2=(k_i+k_{i-1})^2/4 + \Delta_{i\ell}^2/12$
for a non-resonance bin and $\epsilon_{i}=\epsilon_{\rm res}$ for a
resonance one. 
Comparing the approximate form (\ref{appro-expansion}) with
the exact one (\ref{exact-expansion})
in the asymptotic region $R \to \infty$,
it is natural to assume 
\begin{equation}
\label{S-matrix-AV}
S^{(J)}_{\ell,L}(k)
\approx\frac{\hat{S}_{\gamma,\gamma_0}}{W_{\gamma}}
         f_{i\ell}(k)
\quad
 {\mbox{(for Av) ,}}
\end{equation}
for $k$ belonging to the $i$-th bin, i.e., $k_{i-1} < k \le k_i$.

\subsection{The pseudo-state method}
\label{sec:PS-method}

In the PS method, $H_{\rm bc}$ is
diagonalized in a space spanned by a finite number of
$L^2$ type basis functions. The resultant eigenstates
can well reproduce both bound and 
continuous states within a finite region of $k$ and $r$~\cite{YK1}.
The $k$ continuum is automatically discretized by
identifying the eigenstates of positive energies
with $\hat{\Phi}_{i\ell}(r)$.
The weight factor $W_\gamma$ is unity if 
the resultant discretized states $\hat{\Phi}_{i\ell}(r)$ are
orthonormalized.
Among the eigenstates, only low-lying states belonging to the
region $0 < \epsilon < \hbar^{2} k_{\rm max}^2/2\mu_{\rm bc}$ are
taken as breakup channels in CDCC equation (\ref{eqs:CDCC}),
where $k_{\rm max}$ is the maximum $k$ in the Av method.

The CDCC equations (\ref{eqs:CDCC}) thus 
obtained
yield discrete breakup $S$-matrix elements.
If the basis functions form a complete set with good accuracy in 
the region of $r$ and $k$ which is important for breakup processes, 
an accurate transformation from the approximate breakup $S$-matrix
elements to the continuous (``exact'') ones is possible, 
as shown in \cite{Tostevin} for the Av method. 
The exact breakup $T$-matrix element is given by
\begin{eqnarray}
   T^{(J)}_{\ell L}(k) =
\langle \Phi_{\ell}(k,r) j_{L}(PR) {\cal Y}_{JM}^{\ell,L}
| U | \Psi_{JM} \rangle 
\;.
\label{T-matrix}
\end{eqnarray}
Inserting the approximate complete set $\{\hat{\Phi}_{i\ell}(r)\}$ between
the bra vector and the operator $U$ in Eq.~(\ref{T-matrix}),
and replacing the ket vector by
the CDCC wave function (\ref{appro-expansion}), one obtains the following
approximate relation,
\begin{eqnarray}
   T^{(J)}_{\ell L}(k) &\approx& \sum_{i} f^{\rm PS}_{i\ell}(k)
\langle \hat{\Phi}_{i\ell}(r)j_{L}(PR) {\cal Y}_{JM}^{\ell,L}| U
| \Psi^{\rm CDCC}_{JM} \rangle 
 \nonumber \\
&\approx& \sum_{i} f^{\rm PS}_{i\ell}(k) \hat{T}_{\gamma,\gamma_0} ,
\label{T-matrix-appro}
\end{eqnarray}
where
\begin{eqnarray}
\label{overlap-PS}
f^{\rm PS}_{i\ell}(k)&=&\langle\Phi_{\ell}(k,r)|
\hat{\Phi}_{i\ell}(r)\rangle
\end{eqnarray}
 and
\begin{eqnarray}
\hat{T}_{\gamma,\gamma_0}&=&
\langle\hat{\Phi}_{i\ell}\,j_{L}(\hat{P}_{i}R)
{\cal Y}_{JM}^{\ell,L}| U
| \Psi^{\rm CDCC}_{JM} \rangle.
\end{eqnarray}
The last form of Eq.~(\ref{T-matrix-appro}) has been derived
by replacing $P$ by
$\hat{P}_i$ in the spherical Bessel function $j_{L}(PR)$,
which is valid since
the $k$ distribution of $f^{\rm PS}_{i\ell}(k)$ is sharply localized
at $k=\hat{k}_i$.
$\hat{T}_{\gamma,\gamma_0}$ is
a CDCC breakup $T$-matrix element calculated with CDCC.
Since the $\hat{T}_{\gamma,\gamma_0}$ are proportional to
the corresponding $S$-matrix elements $\hat{S}_{\gamma,\gamma_0}$,
\begin{equation}
   S^{(J)}_{\ell L}(k) \approx \sum_{i} f^{\rm PS}_{i\ell}(k)
    \hat{S}_{\gamma,\gamma_0} \; .
\label{S-matrix-appro}
\end{equation}
The {\lq\lq}interpolation formula'' (\ref{S-matrix-appro}) 
for the PS method agrees with the corresponding one in 
\cite{Tostevin} for the Av method. 
Thus, the interpolation formula (\ref{S-matrix-appro}) can
be used for any method of discretization, if the discretized wave
functions constitute an approximate complete set.
The interpolation formula (\ref{S-matrix-appro}) is also independent of
the type of the basis function taken, as obvious from the derivation,
but it is necessary that the basis functions
form a complete set in good approximation in the region of 
$r$ and $k$ which is important for the breakup $T$-matrix elements.
As such basis functions, we here propose two types; 
one is the conventional real-range Gaussian functions
\begin{equation}
\label{RG}
r^{\ell}\exp[-(r/a_j)^2],
\end{equation}
where ${a_{j}}$ ($j = 1$--$n$) are
assumed to increase in a geometric
progression~\cite{K-G}:
\begin{equation}
 a_{j}=a_1 (a_{n}/a_1)^{(j-1)/(n-1)}.
\label{aj}
\end{equation}
The other is an extension of Eq.~(\ref{RG})~\cite{H-Ka-Ki}:
pairs of functions
\begin{subequations}
\label{comp-g}
\begin{equation}
r^{\ell}\exp\left[-\left(r/a_{j}\right)^2\right]
\cos\left(b \left(r/a_{j}\right)^2\right),
\label{comp-g-cos}
\end{equation}
\begin{equation}
r^{\ell}\exp\left[-\left(r/a_{j}\right)^2\right]
\sin\left(b\left(r/a_{j}\right)^2\right),
\label{comp-g-sin}
\end{equation}
\end{subequations}
which can be also expressed
as $(\phi_{i\ell}+\phi_{i\ell}^{*})/2$ and
$(\phi_{i\ell}^{*}-\phi_{i\ell})/(2i)$, respectively, with
\begin{equation}
\label{CG}
\phi_{i\ell}(r) =r^{\ell}\exp[-\eta_{j} r^2],
\;\;\;
\eta_{j}=(1+i\,b)/a_{j}^2,
\end{equation}
i.e., a Gaussian function with a complex range parameter. 
We refer to the new basis
(\ref{comp-g}) as the complex-range Gaussian basis.
In Eqs.~(\ref{comp-g}) and (\ref{CG}) ${a_{j}}$ are the same as in
Eq.~(\ref{aj}); $b$ is a free parameter, in principle, but 
numerical test show that $b =\pi/2$ is recommendable.
It should be noted that the total number of basis
functions is $2n$.
Note also that the complex-range Gaussian basis
agrees with the real-range Gaussian basis when $b=0$.

The complex-range Gaussian basis functions
 are oscillating with $r$.
They are therefore expected to simulate
the oscillating pattern of the continuous breakup state wave functions
better than the real-range Gaussian basis functions do.
Actually, the latter reproduces the continuous state $\Phi_{\ell}(k,r)$
in the region $0 \le k r \lesssim 20$~\cite{YK1}, while
the former does in the even larger region, $0 \le k r \lesssim 35$.

If necessary, one can calculate, with
the complex-range Gaussian basis as well as
the real-range Gaussian basis, all the coupling potentials
analytically by expanding the potential $U$
into a finite number of Gaussian functions.
As far as three-body breakup reactions are concerned,
however the analytic forms of the coupling potentials are not necessarily,
since they can be easily obtained
with numerical integration.
Direct numerical calculation is difficult in the case of
four-body breakup, and the analytic forms of the potentials are very
useful in that case, as shown in section \ref{sec:discussions}.

\begin{table*}[htbp]
\caption{Parameters of the optical potentials for $n$ + $^{58}$Ni
and $p$ + $^{58}$Ni at the half the deuteron incident energy.
We followed the same notation as in Ref.~\protect\cite{BGP}
}
\begin{center}
\begin{tabular}{cccccccccc}
\hline
\hline \\
 system & $V_0$ (MeV)~ & $r_0$ (fm)~ & $a_0$ (fm)~ &
 $W_0$ (MeV)~
 & $r_{\mathrm{W}}$ (fm)~ & $a_{\mathrm{W}}$ (fm)~
 & $W_{\mathrm{D}}$ (MeV)~ & $r_{\mathrm{WD}}$ (fm)~
 &  $a_{\mathrm{WD}}$ (fm) \\
 \\ \hline \\
 $p$ + $^{58}$Ni & 44.921 & 1.17 & 0.750 & 6.10 & 1.32 & 0.534
 & 2.214   & 1.32  & 0.534 \\ \\
 $n$ + $^{58}$Ni & 42.672 & 1.17 & 0.750 & 7.24 & 1.26 & 0.580
 & 2.586  & 1.26  & 0.580 \\ \\ \hline
\end{tabular}
\end{center}
\end{table*}

\section{Numerical test of the pseudo-state method}

In previous CDCC analyses~\cite{Yahiro1,Piya},
calculated elastic and breakup $S$-matrix elements 
were found to converge, for sufficiently large model space.
 In this section, we test the PS method by comparing the calculated
 $S$-matrix elements with those obtained with the Av method that
 converged within the error of 1\% and 
 hence forth called ``exact'' $S$-matrix elements.
The test is made for two systems, $d$+$^{58}$Ni scattering at 80 MeV and
$^{6}$Li+$^{40}$Ca scattering at 156 MeV.

\subsection{$d$+$^{58}$Ni scattering at 80 MeV}

The model space taken in the present CDCC calculations is
$\ell=0, 2$ and $k_{\rm max}=1.3 ~{\rm fm}^{-1}$.
It should be noted that the p-wave ($\ell=1$) breakup is negligible,
because couplings $V_{\gamma\gamma^{'}}(R)$ between odd and even parity 
breakup states contain a small factor,
$U=U_{\rm pA}({\bf r}_{\rm pA}) - U_{\rm nA}({\bf r}_{\rm pA})$, 
in Eq. (\ref{coupling}). 
The choice of the $k_{\rm max}$ is discussed below.
Table I shows the parameters of the potentials used;
the interaction between a nucleon and the target is the nucleon-nucleus
optical potential of Becchetti and Greenlees~\cite{BGP} at half the deuteron
incident energy.
The interaction between proton and neutron is a
one-range Gaussian potential,
$v_{np}=v_{0}\exp [-(r/r_{0})^2 ]$ with $v_{0}=-72.15$ MeV and
$r_{0}=1.484$ fm, which reproduces the radius and the binding energy of
deuteron.

In the Av method the weight function is taken as $f_{i\ell}(k)=1$,
since the projectile (deuteron) has no resonance state.
The model space that gives convergence
within error of 1\% turns
out to be $\Delta_{i\ell}=1.3/30$ fm$^{-1}$, as mentioned above, and
$k_{\rm max}=1.3$ fm$^{-1}$. 
The resulting values of $k_{\rm max}$ and $\Delta_{i\ell}$ are different
from those used in the previous analysis~\cite{Piya};
the main purpose of Ref.~\cite{Piya} was to show that the convergence of
the CDCC solution was obtained within a model space of practical use and
that the converged solution satisfied an appropriate boundary condition.
The model space taken there, $k_{\rm max}=1.0$ fm$^{-1}$ and
$\Delta_{i\ell}=1/8$ fm$^{-1}$, are indeed
enough for the elastic $S$-matrix elements
and the dominant part of the breakup ones with the smaller $k$, therefore
the elastic cross sections and the total breakup cross sections are well
reproduced.
However, the model space is found to be insufficient 
to obtain the {\lq\lq}exact'' $S$-matrix elements
in the high $k$ region around $1.0 ~{\rm fm}^{-1}$,
hence we take here $k_{\rm max}=1.3$ fm$^{-1}$ and
$\Delta_{i\ell}=1.3/30$ fm$^{-1}$ mentioned above.

\begin{figure}[htbp]
 \begin{center}
  \includegraphics[width=0.4\textwidth,clip]{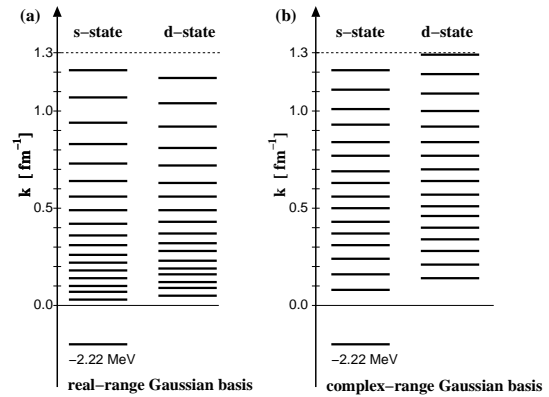}
\caption{Discretized momenta for real-range (a) and complex-range (b)
Gaussian bases for deuteron.
In each panel, the left (right) side corresponds to the
s-state (d-state).
The horizontal dotted line represents the maximum 
momentum $k_{\rm max}$ taken to be 1.3 fm$^{-1}$.}
 \end{center}
\end{figure}

\begin{figure}[htbp]
 \begin{center}
  \includegraphics[width=0.4\textwidth,clip]{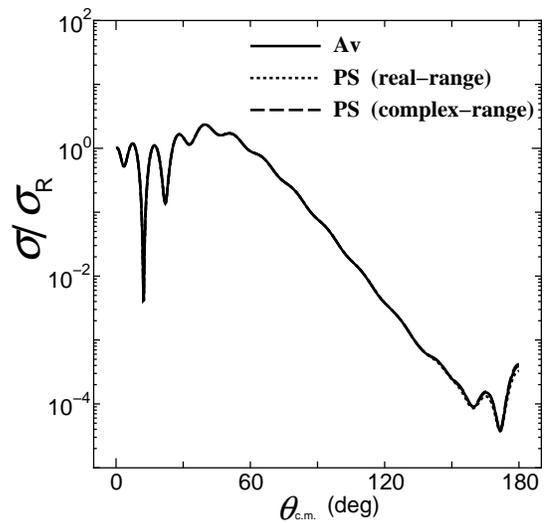}
  \caption{Angular distribution of the elastic differential cross section
  as a ratio to the Rutherford one for $d$ + $^{58}$Ni
  scattering at 80 MeV. Results with the Av method, the real- and
  complex-range Gaussian PS methods are represented by the solid, dashed and
  dotted lines, respectively.}
 \end{center}
\end{figure}

\begin{figure*}[htbp]
 \begin{center}
  \includegraphics[width=0.4\textwidth,clip]{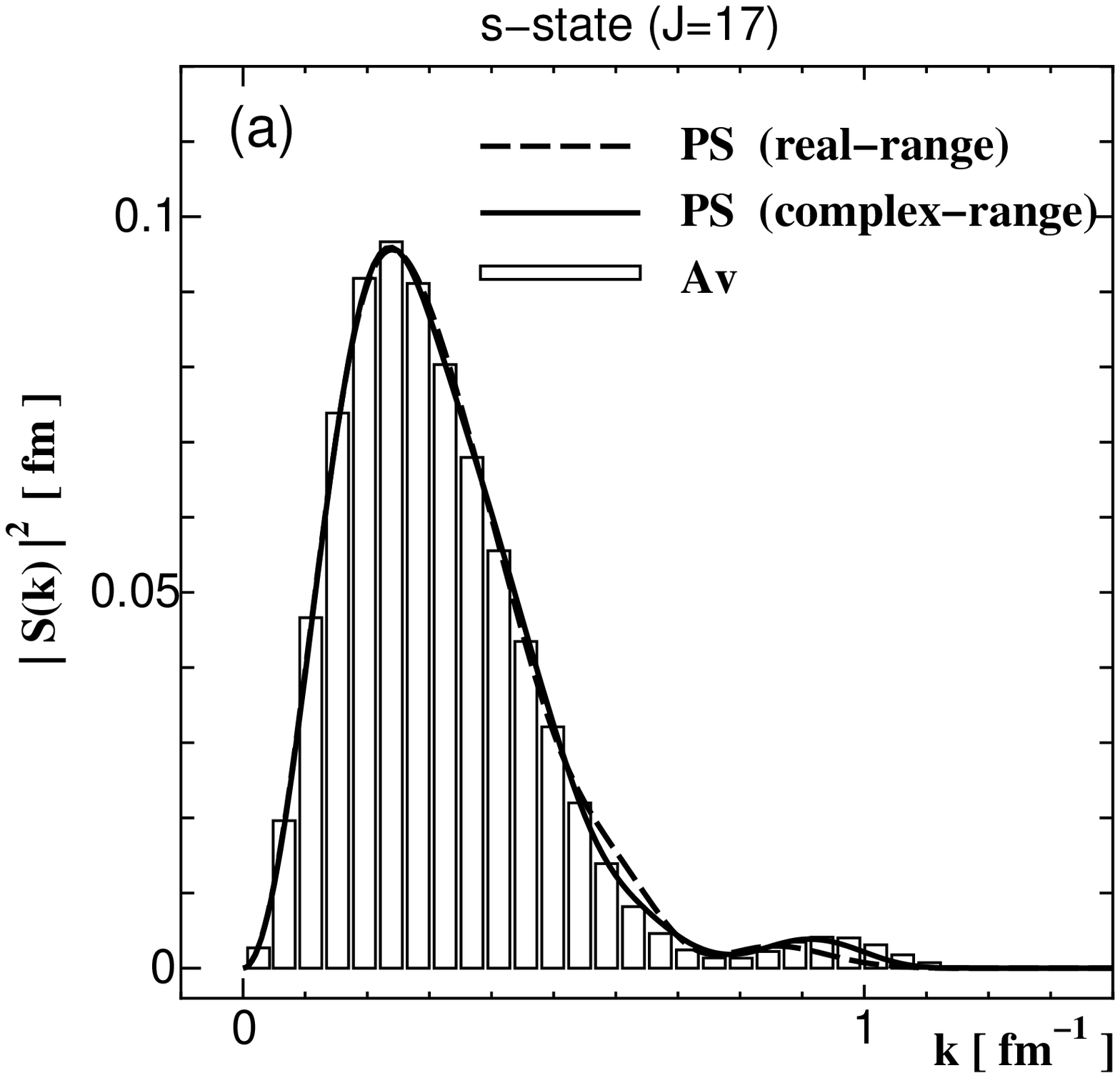}
  \includegraphics[width=0.4\textwidth,clip]{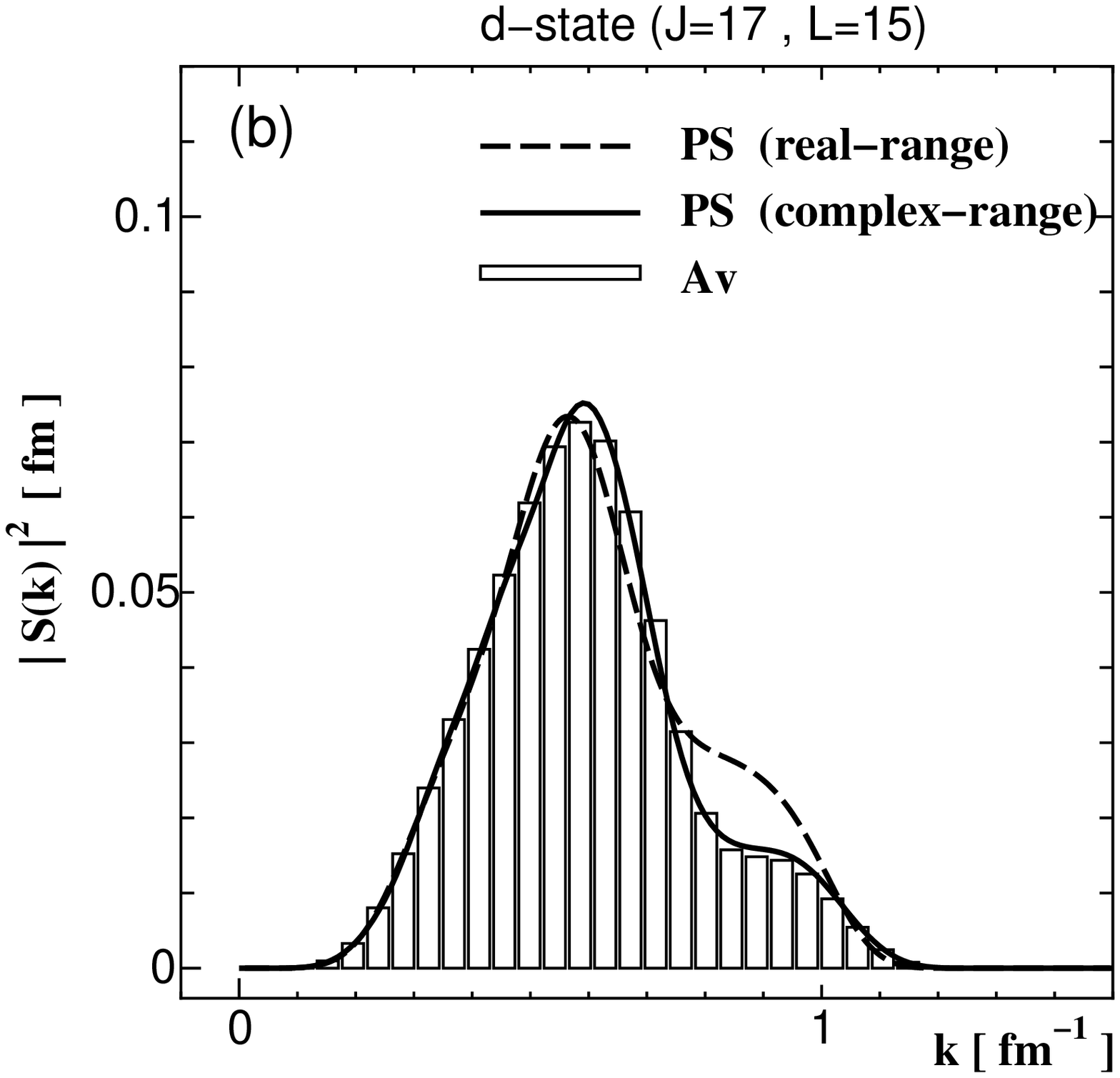}
  \includegraphics[width=0.4\textwidth,clip]{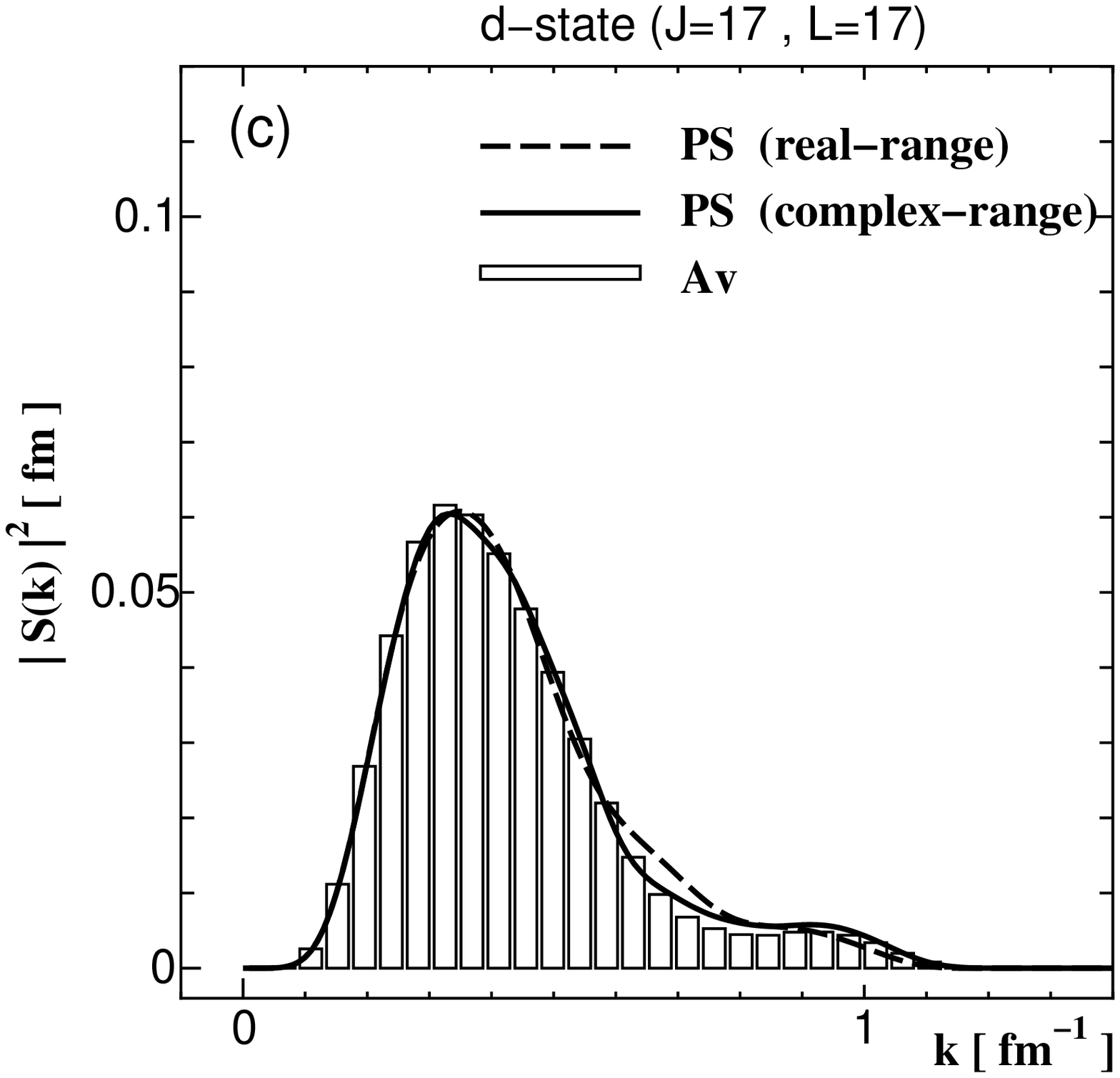}
  \includegraphics[width=0.4\textwidth,clip]{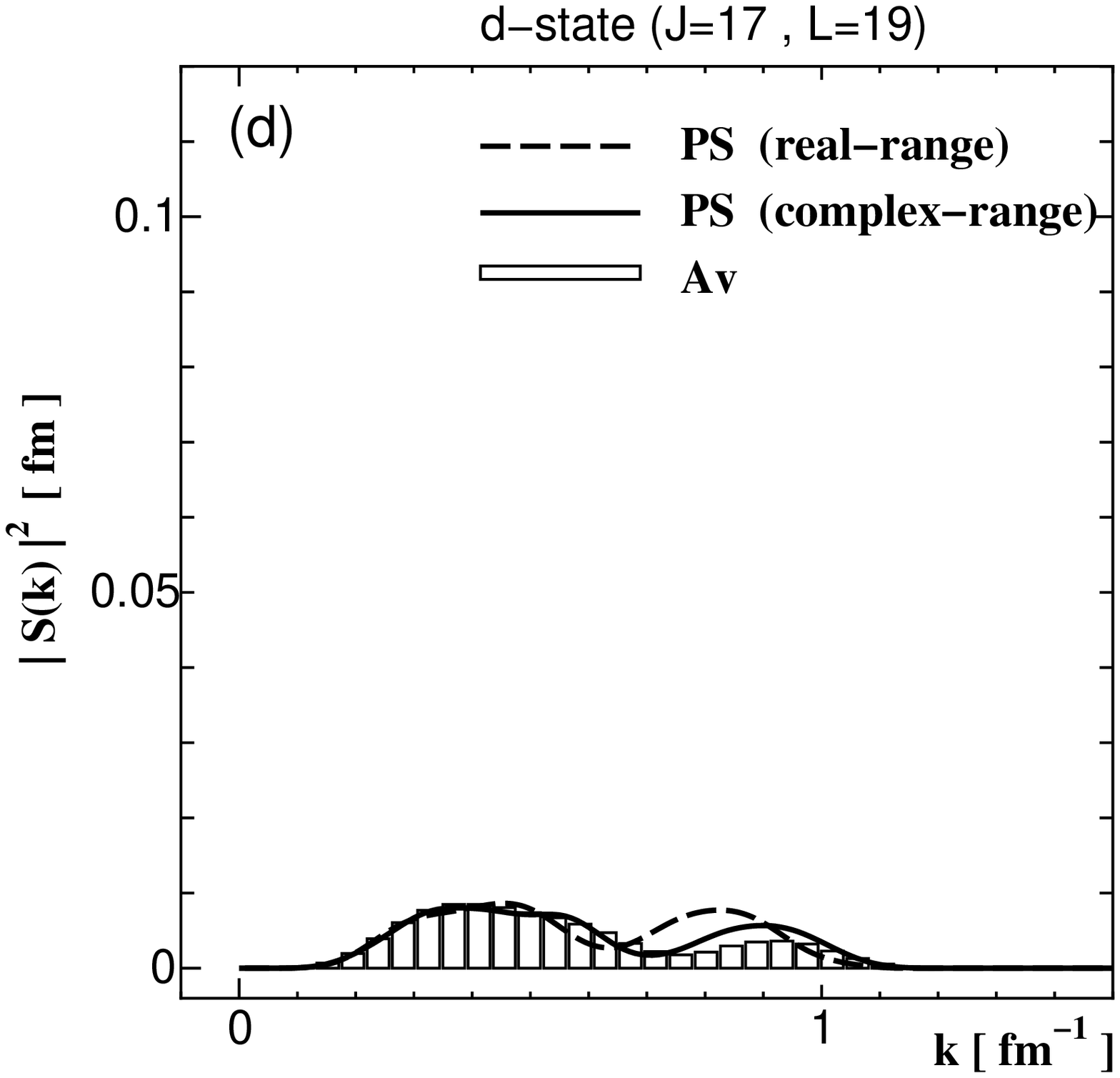}
  \caption{The squared moduli of breakup $S$-matrix elements as a
  function of $k$ at the grazing total angular momentum $J=17$ for $d$ +
  $^{58}$Ni scattering at 80 MeV.
  The panel (a), (b), (c) and (d) shows the result for the s-state with,
  and the d-states with $L=15, 17$ and 19 respectively.
  In each panel, the dashed (solid) line represents the result of the
  real-range (complex-range) Gaussian PS method.
  The step line is the result for the ``exact'' $S$-matrix element
  calculated the Av method.}
 \end{center}
\end{figure*}

In the real-range Gaussian PS method, a similar convergence is found,
when the number of breakup channels, $N_{\rm PS}$, is 18
for both s- and d-waves.
 The number is even smaller when the complex-range Gaussian basis
is taken: $N_{\rm PS}$ is 16 for s-wave and 17 for d-wave.
The basis functions finally obtained have parameter sets
$(a_1=1.0, a_{n}=30.0, n=30 )$
for real-range Gaussian basis and
$(a_1=1.0, a_{n}=20.0, 2n=40, b=\pi/2)$
for complex-range one.
For both of them $N_{\rm PS}$ is smaller than
the number of basis functions. 
High-lying states with $k > k_{\rm max}$, which
are obtained by diagonalizing $H_{\rm bc}$, do not
affect the breakup $S$-matrix elements with $k < k_{\rm max}$, because
the coupling potentials between the two $k$ regions are weak.

Figure 2 shows the discrete momenta $\hat{k}_{i\ell}$
translated from the eigenenergies $\epsilon_{i\ell}$
for the real- and complex-range Gaussian bases. 
One sees that for the real-range Gaussian basis,
the discrete momenta are dense in the smaller $k$ region
and sparse in the larger $k$ one. This distribution is not so effective
in simulating the $k$ continuum, in the higher $k$ region in particular.
For the complex-range Gaussian basis, on the other hand,
the discrete momenta are distributed almost evenly.
A similar sequence of the $\hat {k}_i$ is also seen for the case of
the transformed harmonic oscillator basis of Ref.~\cite{PS1,Perez}.
Such a sequence of $\hat {k}_i$ is close to that in the Av method.
Thus, the complex-range Gaussian basis,
as well as the transformed harmonic oscillator, is well suited for
simulating the $k$ continuum in the entire region $0 < k < k_{\rm max}$.

For the elastic $S$-matrix elements, both  the real- and
complex-range Gaussian PS methods well reproduce the ``exact'' one
calculated 
with the Av method, as confirmed in Fig.~3 for the differential
cross section. The three types of calculations,
the real-range Gaussian PS, the complex-range Gaussian PS
and the Av methods, yield an identical cross section at all scattering
angles. Thus, both of the PS methods proposed here are useful for treating
the breakup effects on the elastic scattering.

Figure 4 shows the result for breakup $S$-matrix elements
at the grazing total angular momentum $J=17$,
as a function of $k$.
The real-range Gaussian PS method (dashed line) well simulates the exact
solution calculated with the Av method (step line)
in the lower $k$ region that corresponds to the main components
of the breakup $S$-matrix elements,
but inaccurate in the higher $k$ region around $k=0.8$ fm$^{-1}$.
The deviation at higher $k$ stems from the fact
that real-range Gaussian basis poorly reproduces the continuum breakup
state $\Phi_{\ell}(k,r)$ at the higher $k$.
Figure 4 shows that this problem can be solved by using the complex-range
Gaussian basis (solid line) instead.

\begin{figure}[htbp]
 \begin{center}
  \includegraphics[width=0.4\textwidth,clip]{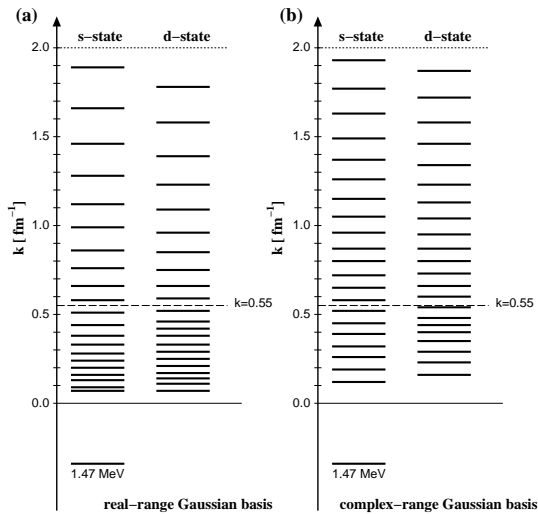}
 \caption{The same as in Fig.~3 but for $^6$Li;
 $k_{\rm max}$ is taken to be 2.0 fm$^{-1}$.
The horizontal dashed line corresponds to the border momentum
between the resonance and non-resonance parts used in the Av method.}
 \end{center}
\end{figure}

\begin{figure}[htbp]
 \begin{center}
  \includegraphics[width=0.4\textwidth,clip]{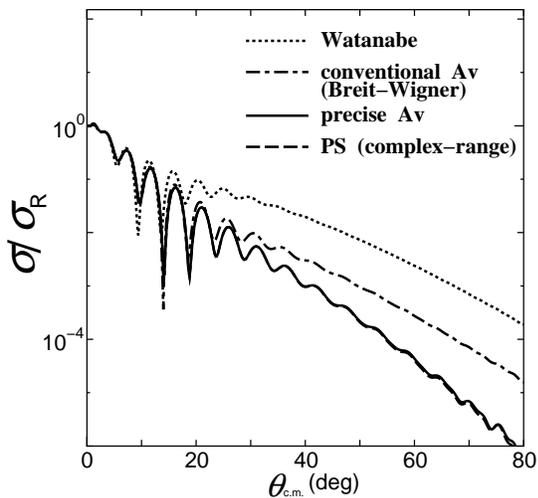}
  \caption{Angular distribution of the elastic differential cross section
  (Rutherford ratio) for $^{6}$Li + $^{40}$Ca scattering at 156 MeV.
  The results of the complex-range Gaussian PS method and
  the approximate treatment of the resonance of $^6$Li,
  i.e., the conventional Av method with the weight factor of Breit-Wigner
  type, are shown by the dashed and dash-dotted lines, respectively.
  The solid line is the ``exact'' solution calculated by the Av method
  with 
  dense bins and the dotted line is the result of Watanabe model, i.e.,
  without breakup effects.}
 \end{center}
\end{figure}

\subsection{$^{6}$Li+$^{40}$Ca scattering at 156 MeV}

\begin{table*}[htbp]
\caption{The same as in Table I but for
$\alpha$ + $^{40}$Ca at 104 MeV and $d$ + $^{58}$Ca at 56 MeV.
}
\begin{center}
\begin{tabular}{ccccccccccc}
\hline
\hline \\
 system & $V_0$ (MeV)~ & $r_0$ (fm)~ & $a_0$ (fm)~ &
 $W_0$ (MeV)~
 & $r_{\mathrm{W}}$ (fm)~ & $a_{\mathrm{W}}$ (fm)~
 & $W_{\mathrm{D}}$ (MeV)~ & $r_{\mathrm{WD}}$ (fm)~
 &  $a_{\mathrm{WD}}$ (fm) \\
 \\ \hline \\
 $\alpha$ + $^{40}$Ca & 219.30 & 1.21 & 0.713 & 98.8 & 1.40 &0.544
 &- & - & - \\ \\
 $d$ + $^{40}$Ca & 75.470 & 1.20 & 0.769 & 2.452 & 1.32 & 0.783
 & 9.775 & 1.32 & 0.783  \\ \\ \hline
\end{tabular}
\end{center}
\end{table*}

Characteristic to this scattering,
the projectile ($^{6}$Li) has d-wave triplet resonance states
($3^+, 2^+, 1^+$). 
For simplicity, we neglect the intrinsic spin of
$^{6}$Li, following Refs.~\cite{CDCC-review1,sakuragi1,sakuragi2}. 
Then the projectile has only one d-wave resonance state 
with $\epsilon_{\rm res}=2.96$~MeV and $\Gamma=0.62$~MeV. 
Obviously the energy and the width do not reproduce experimental data, 
but at least the elastic cross section of $^{6}$Li 
is not affected much by the neglection of the spin \cite{Tanifuji}.

In this scattering, the three-body system consists of deuteron,
$\alpha$ and $^{40}$Ca. The interactions between each pair of the
constituents are the optical potential of
$\alpha+^{40}$Ca scattering at 104 MeV~\cite{HAU72},
that of $d+^{40}$Ca scattering at 56 MeV~\cite{dCa},
and $v_{\rm \alpha d}=v_{0}\exp\left[-(r/r_{0})^2\right]$
with $v_{0}=- 74.19$~MeV and $r_{0}=2.236$~fm.
Table II shows the parameters of the optical potentials.

\begin{figure*}[htbp]
 \begin{center}
  \includegraphics[width=0.4\textwidth,clip]{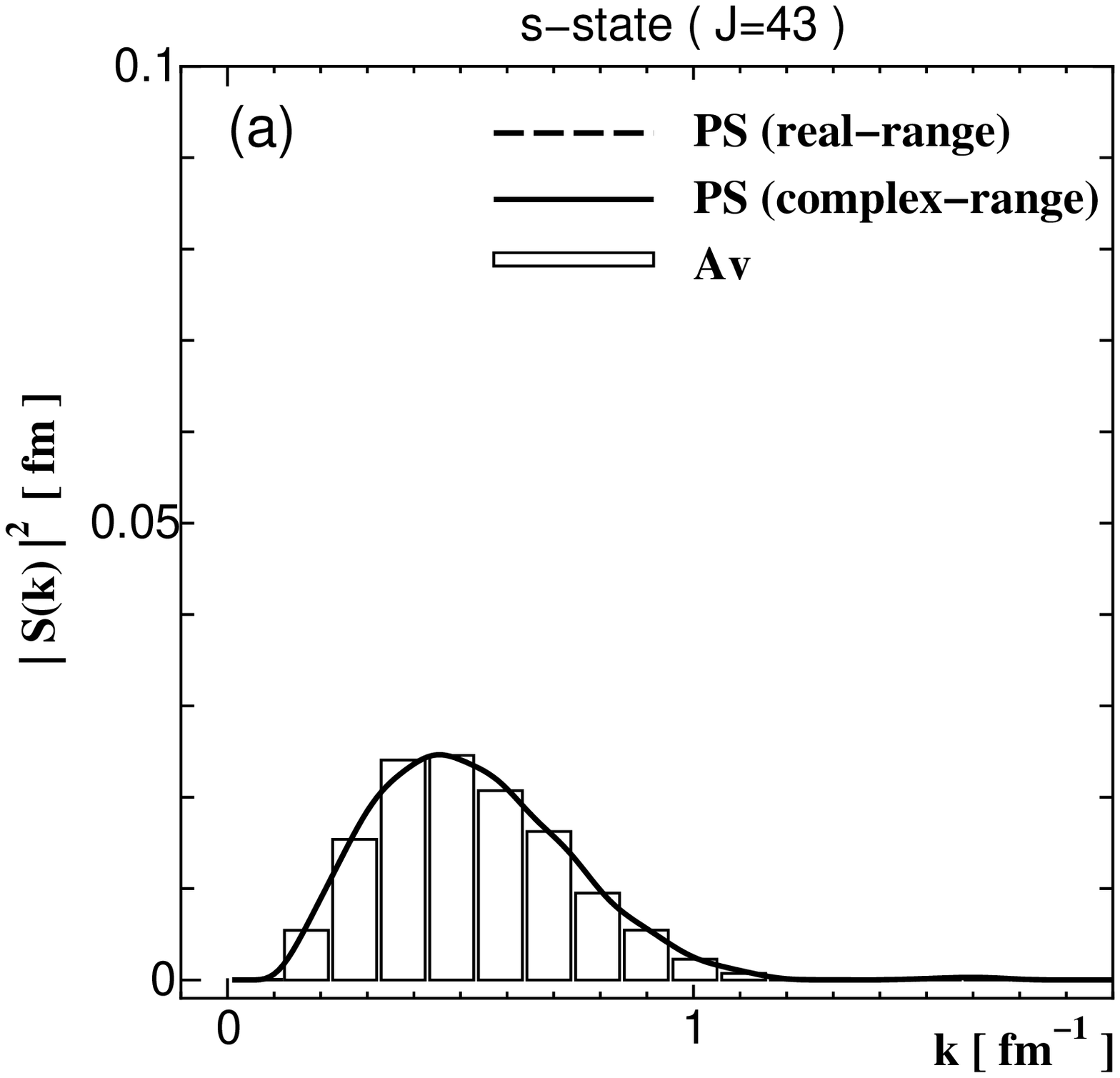}
  \includegraphics[width=0.4\textwidth,clip]{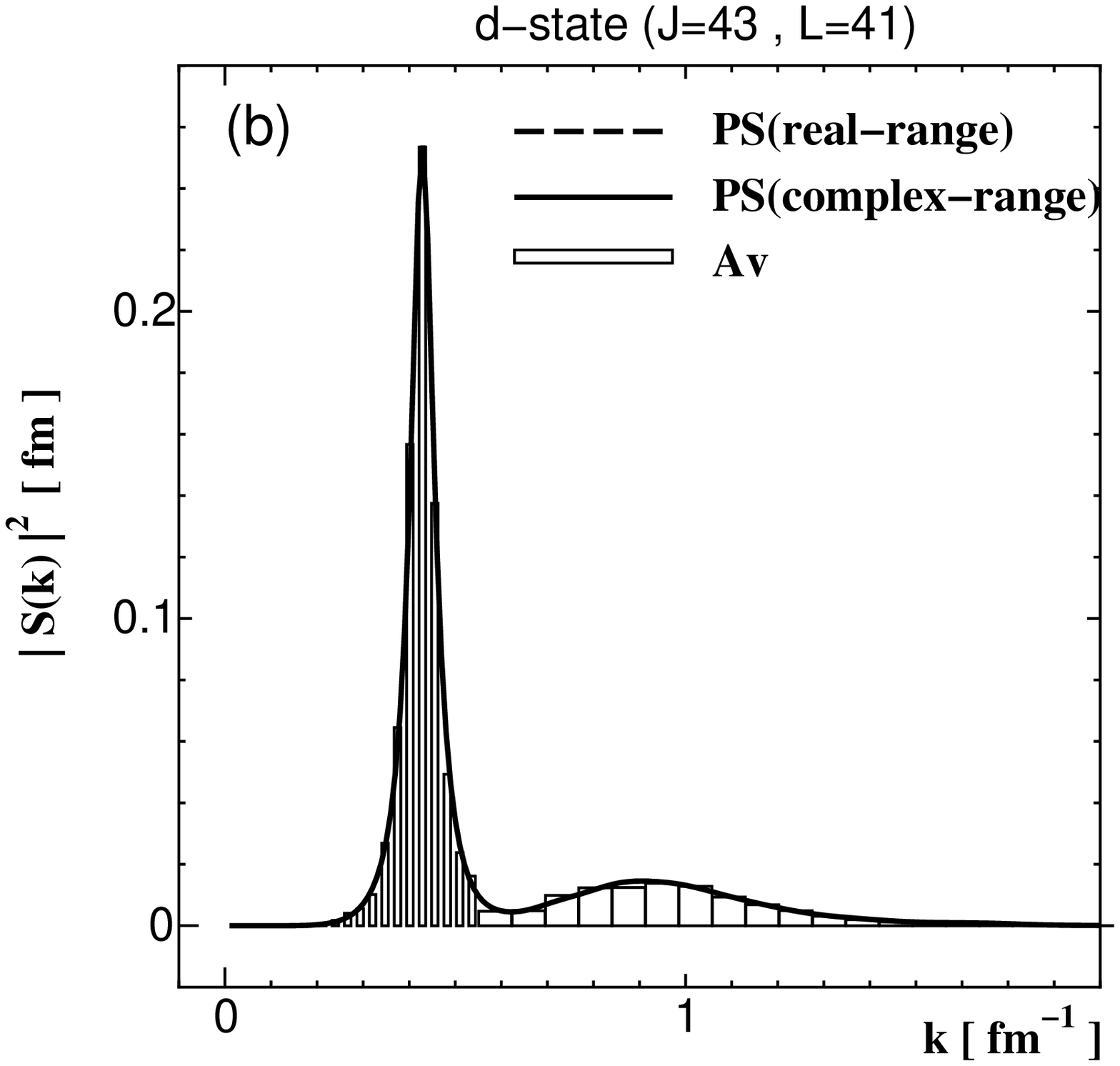}
  \includegraphics[width=0.4\textwidth,clip]{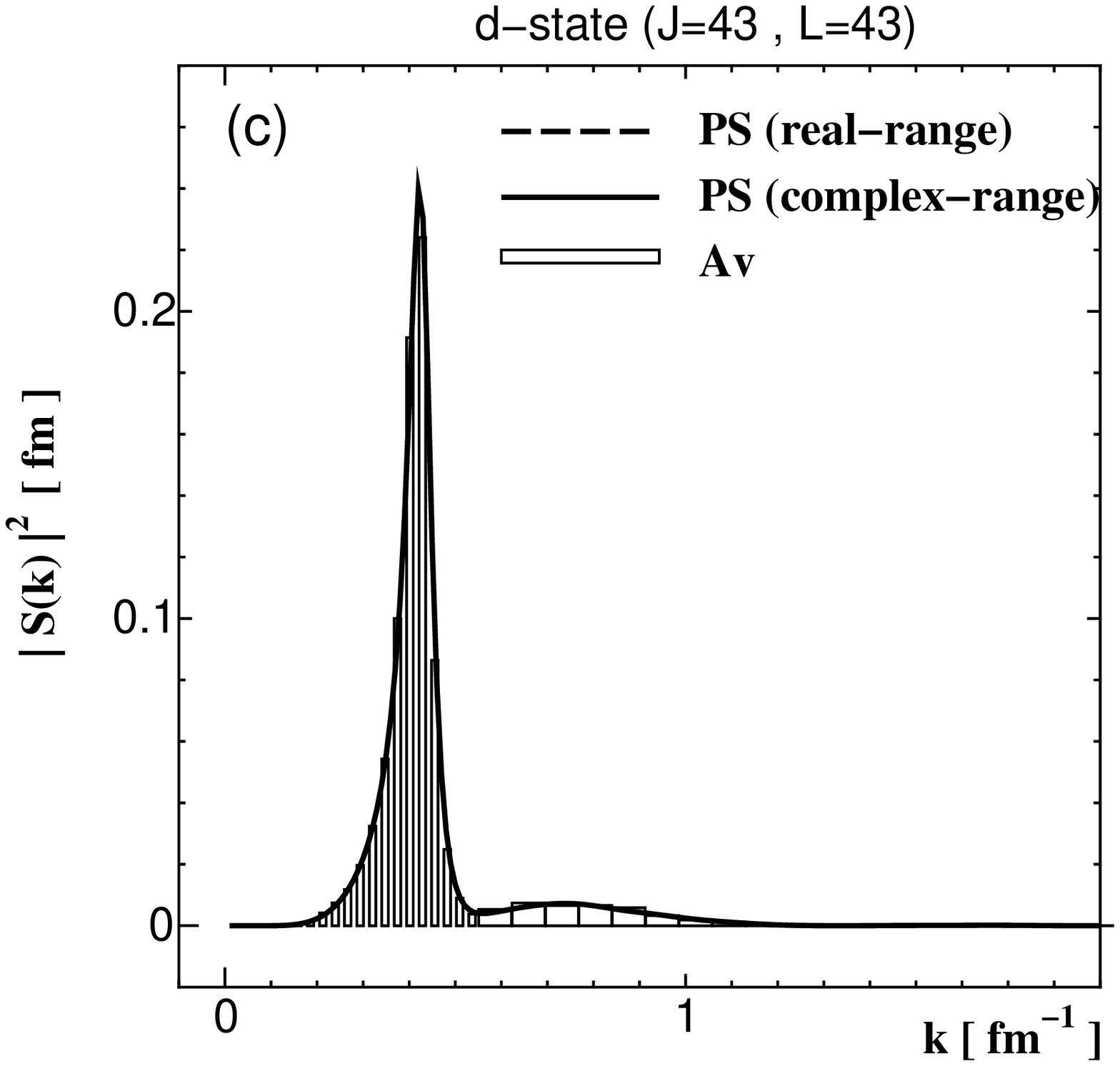}
  \includegraphics[width=0.4\textwidth,clip]{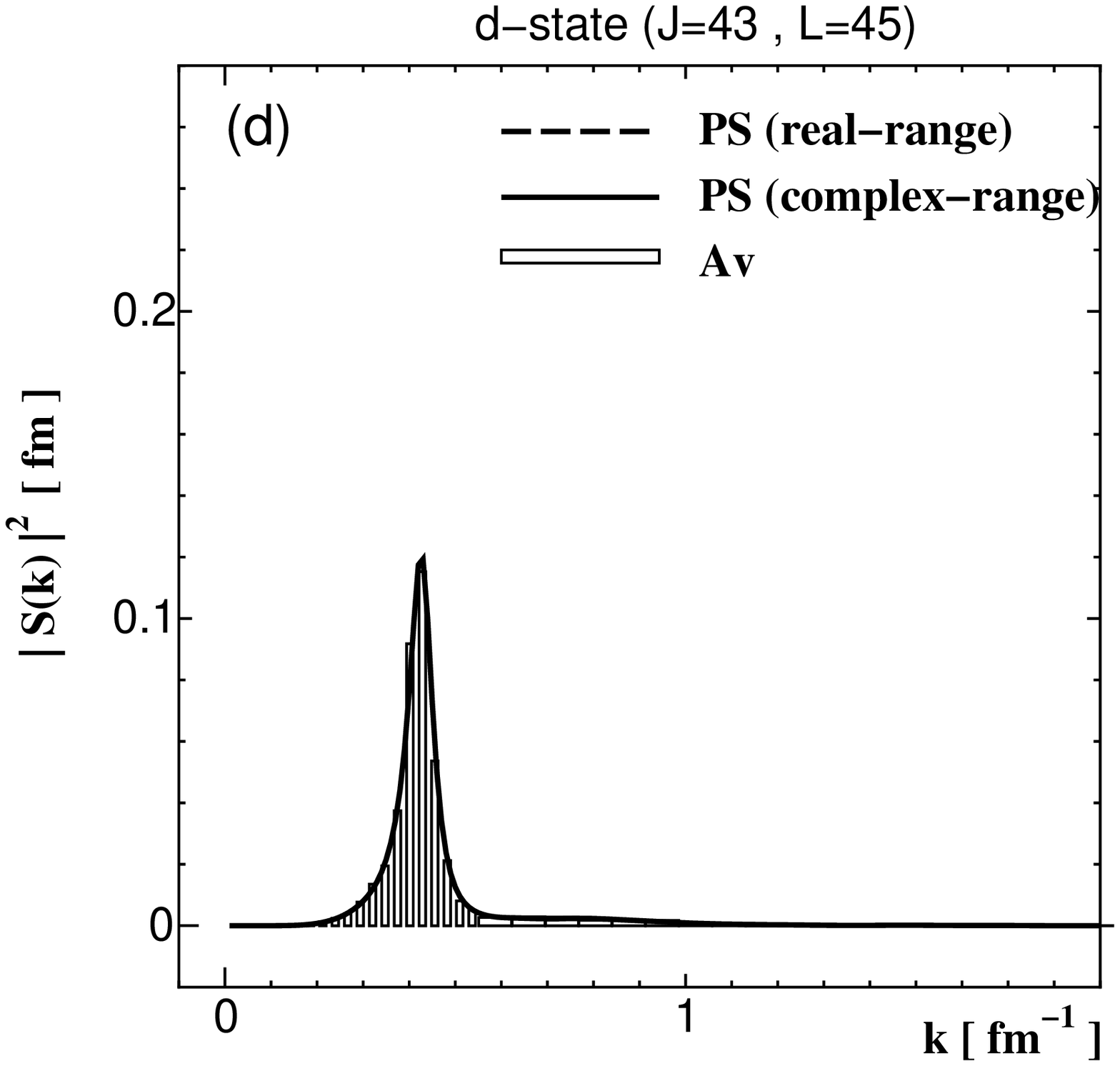}
  \caption{The same as in Fig.~4 but for $^6$Li + $^{40}$Ca scattering
  at 156 MeV. The corresponding grazing total angular momentum is 43.
  The step line is the result of the Av method with dense bins. Note that
  the difference between the results of the real- and complex-range
  Gaussian PS methods is not visible since it is less than
  about 1\%.}
 \end{center}
\end{figure*}

The model space sufficient for describing breakup processes in this
scattering 
is $k_{\rm max}=2.0$ fm$^{-1}$ and $\ell_{\rm max}=2$;
the model space is composed of two $k$-continua for $\ell=0$ and 2.
Since there exists a resonance in $\ell=2$, the
d-wave $k$-continuum is further divided in the Av method into
the resonant part $[0 < k < 0.55]$ and the non-resonant part
$[0.55 < k <2.0]$.
The $k$ continuum of $\Phi_{i,\ell=2}(k,r)$ in the resonant part
varies rapidly with $k$.
The Av method can simulate the rapid change
by taking $f_{i,\ell=2}(k)=1$
with bins of an extremely small width.
In fact clear convergence is found for both the elastic and the breakup 
$S$-matrix elements, when the resonance part is described by 30 bins and
the non-resonance part of the d-wave and the s-wave $k$-continua
by 20 bins.
Another Av discretization, which has been widely used as a convenient
prescription~\cite{CDCC-review1,sakuragi1,sakuragi2,sakuragi4,Sakuragi},
is also made for comparison, in which the resonance region is
represented by a single state with the weight factor of
Breit-Wigner type (\ref{BW-weight}).
The two sorts of Av discretization are compared
with the real- and complex-range Gaussian PS methods.
With the PS methods,
convergence of the $S$-matrix elements is found
with 21 s-wave breakup channels and 22 d-wave ones.
The level sequence of the resulting discrete eigenstates is shown
in Fig.~5 for both the basis functions. The level sequences have
the same properties as in Fig.~2.
The parameter sets of the basis functions, finally taken in the PS
methods, 
are $(a_1=1.0, a_{n}=20.0, 2n=40, b=\pi/2)$ for the complex-range Gaussian
basis and $(a_1=1.0, a_{n}=30.0, n=30)$ for the real-range one.

Figure 6 shows the differential cross section of
the elastic scattering.
The result with the precise Av discretization based on dense bins,
considered to be the ``exact'' solution, is denoted by the solid line.
The dotted line represents the result of the Watanabe model, i.e.,
with no breakup channels.
The conventional Av discretization, based on the weight factor of
Breit-Wigner type (dash-dotted line),
well describes the breakup effect, particularly
at very forward angles ($\theta < 20^\circ$),
but deviates considerably from the ``exact'' solution
at larger angles ($\theta > 30^\circ$). 
The complex-range Gaussian PS discretization
(dashed line) well reproduces the ``exact'' solution with a number of
channels being suitable for practical use.
The real-range Gaussian PS method gives just the same
result as the complex-range one.

Figure 7 represents breakup $S$-matrix elements
at grazing total angular momentum $J=43$.
The real- and complex-range Gaussian PS discretization well reproduce
the ``exact'' solution calculated by the Av discretization with dense
bins. 
The results of the two PS methods turn out to coincide within the
thickness of the line.
The resonance peak can be expressed by only 8 (12) breakup
channels in the complex-range (real-range) Gaussian PS method,
while the corresponding number of breakup channels is 30 in the Av
method, as mentioned above.
Thus, one can conclude that the real- and complex-range Gaussian PS
methods are very useful for describing not only non-resonant states but
also resonant ones.

\section{Discussions on Four-body breakup reaction}
\label{sec:discussions}

In the past CDCC calculations the projectile was assumed to be a
two-body system, 
dealing only with three-body breakup reactions.
In this section, we investigate the applicability of CDCC to four-body
breakup reactions of the projectile consisting of three particles, b+c+x
(Fig.~8).
The Av method needs the exact three-body wave functions
being impossible to obtain.
We can circumvent this problem with the present PS method;
one can prepare an approximate complete set
$\{ \hat{\Phi}_{i\ell} \}$ by diagonalizing the Hamiltonian of
the projectile in a space spanned by a set of basis functions of $L^2$
type. With $\{ \hat{\Phi}_{i\ell} \}$ as the wave functions of the
breakup channels, one can obtain an approximate total wave function
$\Psi^{\rm CDCC}$ by solving CDCC equations (\ref{eqs:CDCC}). 
Inserting $\Psi^{\rm CDCC}$ into the exact form of breakup $T$-matrix
elements in place of the exact total wave function, 
one reaches an approximate form:
\begin{eqnarray}
   T_4 =
\langle
 e^{ i({\bf P} \cdot {\bf R} + {\bf k} \cdot {\bf r} + {\bf q}
\cdot {\bf y})}|
U_{4}|\Psi^{\rm CDCC} \rangle_{{\bf R,r,y}},  
\label{T4}
\end{eqnarray}
where $U_{4}$ is the sum of all interactions in the four-body 
system (A+b+c+x), 
${\bf r}$ and ${\bf y}$ are and ${\bf k}$ and ${\bf q}$ are,
respectively, two Jacobi coordinates of the three-body (b+c+x) system 
and momenta conjugate to them. 
The accuracy of Eq.~(\ref{T4}) depends on how complete the 
set $\hat{\Phi}_{i\ell}$ is within the region
$(0 \le r \le r_{\rm max},0 \le y \le y_{\rm max})$
which is important for the breakup process considered.
An important advantage of the use of the real- and complex-range
Gaussian bases is, as mentioned before, that analytic integrations over
${\bf r}$ and ${\bf y}$ can be done in Eq.~(\ref{T4}),
by expanding $U_4$ in terms of a finite number of Gaussian basis
functions. This makes the derivation of $T_4$ feasible. Analyses based
on this formulation are of much interest as a future work.

\begin{figure}[htbp]
 \begin{center}
  \includegraphics[width=0.3\textwidth,clip]{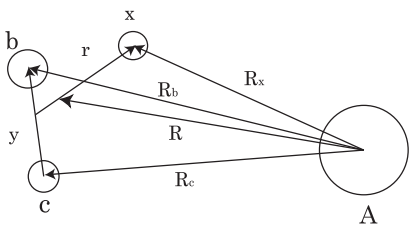}
  \caption{Illustration of a four-body (A+b+c+x) system. The projectile
  consists of b, c and x, and A is the target.}
 \end{center}
\end{figure}

\section{Summary}

The method of continuum discretized coupled channels (CDCC)
is an accurate method of treating three-body breakup
processes, in which the discretization of the $k$ continuum is essential.
In this paper, we propose the new method of pseudo-state (PS)
discretization which can be used not only for virtual breakup processes
in elastic scattering but also for breakup reactions. 
First we show that an accurate transformation from 
the discrete breakup S-matrix elements calculated with the PS method 
to continuous ones is possible, 
since the PS basis functions can form in the good 
approximation a complete set  in the finite region of 
${\bf r}$ and ${\bf k}$ which is important for the breakup processes.
As bases satisfying 
the completeness, 
we propose the real- and complex-range Gaussian bases. 
Both bases can treat virtual breakup processes
in the elastic scattering with high accuracy, 
i.e., with the error of calculated cross sections less than 1\%.
For breakup processes, the complex-range Gaussian basis
is accurate throughout the entire region of the $k$ continuum.
The real-range Gaussian basis also keeps a good accuracy
for the dominant part of breakup $S$-matrix elements with the lower $k$,
although it is partially inaccurate for the higher $k$ region.
Thus, both bases can be used for realistic analyses
of elastic scattering and projectile breakup reactions.

The present new PS method 
has at least two advantages over the 
widely used momentum bin average method. 
One is that it does not need the exact wave function of the projectile
over the entire region of $r$. This is important from a theoretical point
of view.
The other advantage is that with the real- and complex-range
Gaussian bases one can calculate all the coupling potentials
semi-analytically, which is very useful in actual calculations.
Furthermore, if the projectile  has resonances in its excitation spectrum,
the new method discretizes the complicated spectrum 
with a reasonable number of the basis functions,
without distinguishing the resonance states from non-resonant continuous
states. 
These advantages of the new method are extremely helpful, sometimes even
essential, in applying CDCC to four-body breakup effects of
unstable nuclei such as $^6$He and $^{11}$Li.
Actually, a CDCC analysis of four-body breakup effect on the $^6$He
elastic scattering is in progress, and the result of the analysis will
appear in a forthcoming paper.

\section*{Acknowledgments}
The authors would like to thank M. Kawai for helpful discussions.
This work has been supported in part by the Grants-in-Aid for
Scientific Research (12047233, 14540271)
of Monbukagakusyou of Japan.


\end{document}